\documentclass{WileyMSP-template}

\usepackage{graphicx}
\usepackage{bm}
\usepackage{amsmath}
\usepackage{amssymb}
\usepackage{xr}
\usepackage{xcolor}

\begin{document}

\pagestyle{fancy}
\rhead{\includegraphics[width=2.5cm]{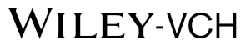}}

\title{Chemical design rules for non-fullerene acceptors in organic solar cells}

\maketitle

\author{Anastasia Markina}
\author{Kun-Han Lin}
\author{Wenlan Liu}
\author{Carl Poelking}
\author{Yuliar Firdaus} 
\author{Diego Rosas Villalva}
\author{Jafar~I.~Khan}
\author{Sri H. K. Paleti}
\author{George T. Harrison} 
\author{Julien Gorenflot}
\author{Weimin Zhang}
\author{Stefaan De Wolf}
\author{Iain~McCulloch}
\author{Thomas D. Anthopoulos}
\author{Derya Baran}
\author{Fr\'ed\'eric Laquai}
\author{Denis Andrienko*}


\begin{affiliations}
Dr. Anastasia Markina, Dr. Kun-Han Lin, Dr. Wenlan Liu, Dr. Carl Poelking, Dr. Denis Andrienko\\
Address: Max Planck Institute for Polymer Research, Ackermannweg 10, 55128 Mainz, Germany\\
Email Address: denis.andrienko@mpip-mainz.mpg.de\\ \qquad
\\
Dr. Yuliar Firdaus, Diego Rosas Villalva, Dr. Jafar~I.~Khan, Sri H. K. Paleti, Dr. George T. Harrison, Dr. Julien Gorenflot, Dr. Weimin Zhang, Prof. Stefaan De Wolf, Prof. Iain~McCulloch, Prof. Thomas D. Anthopoulos, Prof. Derya Baran, Prof. Fr\'ed\'eric Laquai\\
Address: King Abdullah University of Science and Technology (KAUST), Thuwal, Kingdom of Saudi Arabia
\end{affiliations}



\begin{abstract}

Efficiencies of organic solar cells have practically doubled since the development of non-fullerene acceptors (NFAs). However, generic chemical design rules for donor-NFA combinations are still needed. We propose such rules by analyzing inhomogeneous electrostatic fields at the donor-acceptor interface. We show that an acceptor-donor-acceptor molecular architecture, and molecular alignment parallel to the interface, result in energy level bending that destabilizes the charge transfer state, thus promoting its dissociation into free charges. By analyzing a series of PCE10:NFA solar cells, with NFAs including Y6, IEICO, and ITIC, as well as their halogenated derivatives, we suggest that the molecular quadrupole moment of ca 75 Debye {\AA} balances the losses in the open circuit voltage and gains in charge generation efficiency.

\end{abstract}


\section{Introduction}

Replacing fullerene acceptors with strongly-absorbing dyes led to an approximately two-fold increase of the power conversion efficiency (PCE) of organic solar cells (OSCs)~\cite{armin_history_2021, meredith_nonfullerene_2020}. At present, OSCs based on small molecule non-fullerene acceptors (NFAs), blended with donor polymers, have certified power conversion efficiencies up to 17.9~\% for single junctions~\cite{cui_over_2019, xu_single-junction_2019, lin_171_2020, liu_18_2020, li_non-fullerene_2021, qin_chlorinated_2021, zhan_layer-by-layer_2021, jin_d18_2021} and 18.6~\% for all-organic solution-processed tandem cells~\cite{meng_organic_2018, salim_organic_2020}, while fullerene-based OSCs are only 10~\% efficient~\cite{zhao_efficient_2016}. To achieve this milestone, various design strategies have been explored, e.g., modification/manipulation of the Y6 acceptor side chain design~\cite{li_non-fullerene_2021}, the use of ternary mixtures with a vertical phase distribution~\cite{zhan_layer-by-layer_2021}, the chemical modification via chlorination~\cite{qin_chlorinated_2021} or a variation of a fused-ring acceptor block of the donor polymer~\cite{jin_d18_2021}. 

Currently, NFAs match their inorganic counterparts in terms of current generation, but are lacking with regard to their open circuit voltage~\cite{green_solar_2020}. Efficiency losses can be traced back to energy losses during the photon to free charge conversion, and are in general lower than in the fullerene-based cells~\cite{xu_efficient_2020, tang_benzotriazole-based_2019, an_solution-processed_2020}.

Free charge generation in organic solar cells is comprised of two steps. During the first step, a photogenerated exciton dissociates at the donor-acceptor interface into an interfacial charge transfer (CT) state. During this process, the ionization energy or electron affinity offset at the heterojunction provides the driving force for the hole or electron transfer. It is known that this offset should exceed a threshold value in order to enable efficient dissociation of the excited state~\cite{dimitrov_energetic_2012, hendriks_dichotomous_2016,karuthedath_intrinsic_2021}. For NFAs, only ionization energy offsets are relevant, because of the fast energy transfer from donors to acceptors~\cite{karuthedath_intrinsic_2021}.

During the second step of charge separation, the interfacial CT state dissociates into a pair of free charges, or the charge separated (CS) state. This dissociation is expected to be an endothermic process, and the exact mechanism behind the driving force for this process is still under debate~\cite{shen_hot_2015, shoaee_charge_2013, burke_beyond_2015, vandewal_efficient_2014, benduhn_intrinsic_2017, nakano_anatomy_2019}. It is, however, one of the key processes in OSCs, since the energetics and dynamics of the dissociating CT state determines the open circuit voltage of organic heterojunctions~\cite{benduhn_intrinsic_2017, benduhn_impact_2018, vandewal_cost_2020, sini_molecular_2018}.

Both steps involved in the free charge generation can be optimized by an appropriate design of the donor-acceptor pair. The main difficulty in formulating generic chemical design rules for OSC materials is that any changes to the chemical structure simultaneously modify the open-circuit voltage, $V_\text{oc}$, the short-circuit current, $J_\text{sc}$, and the fill factor of the solar cell~\cite{albrecht_efficiency_2014, bartesaghi_competition_2015, tietze_correlation_2013, graham_roles_2016, hou_organic_2018, alamoudi_impact_2018}. Without knowing how these changes correlate with each other, it is impossible to formulate clear design rules and hence speed up the discovery of efficient donor-acceptor combinations.

\begin{figure}[ht]
        \centering
	\includegraphics[width=0.6\linewidth]{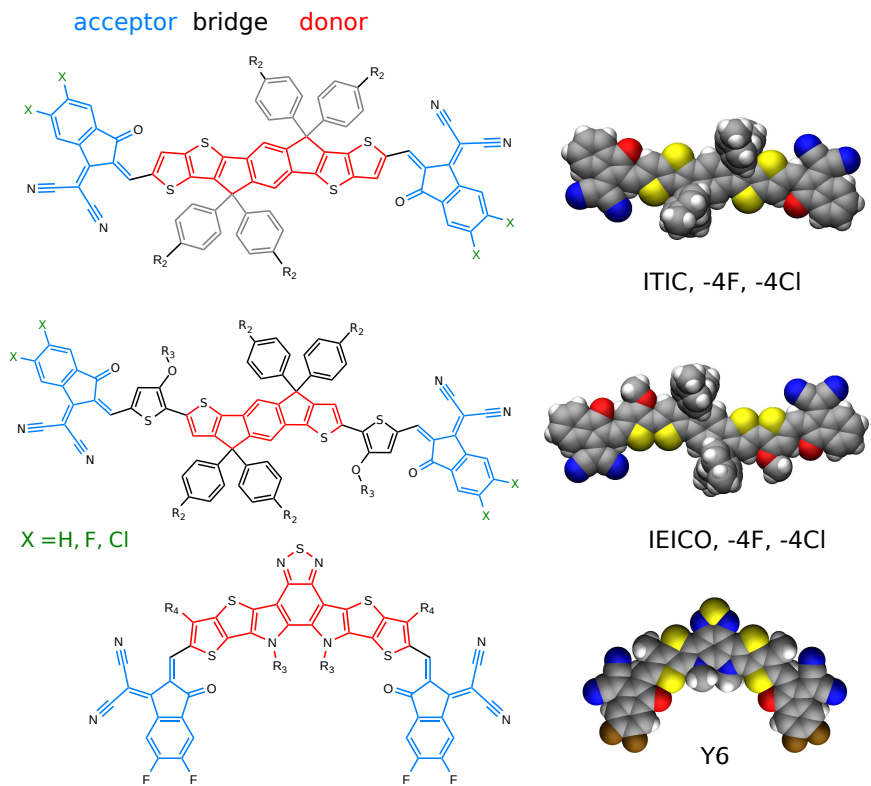}
	\caption{
	Chemical structures of NFAs studied here: IEICO, IEICO-4F, IEICO-4Cl, ITIC, ITIC-4F, ITIC-4Cl, and Y6~\cite{baran_reducing_2017, lin_electron_2015, perdigon-toro_barrierless_2020, wang_ultra-narrow_2018, yu_recent_2018, yang_effects_2019}.
	Color coding emphasizes the acceptor-bridge-donor-bridge-acceptor design patterns. The van der Waals surfaces of optimized conjugated cores illustrate elongated molecular shapes and planarity. Side chains are omitted for clarity. The block-like design simplifies the chemical tuning of optical, transport, and electrostatic properties of NFAs.
	}
	\label{fig:structures}
\end{figure}

In this work, we identify the microscopic origin of such correlations and propose clear chemical design rules for NFAs. To do this, we first evaluate the electrostatic potential at the donor-acceptor interface and show that it can lead to either stabilization or destabilization of the CT state, depending on the quadrupole moment of a neutral NFA molecule, and molecular orientations at the donor-acceptor interface. We then show that the acceptor-donor-acceptor (A-D-A) molecular architecture, present in all efficient NFAs, is effectively responsible for the CT state destabilization. Finally, we provide explicit links between OSC and molecular properties, such as gas-phase ionization energy, electron affinity, quadrupole moment, solid-state crystal fields, and interfacial disorder. All correlations are illustrated on experimentally characterized donor-NFA combinations, which include seven NFAs, as shown in {\bf Figure~\ref{fig:structures}}, combined with a polymer donor, PCE10 (PTB7-Th).

\section{Energy level bending and interfacial bias}

\begin{figure}[ht]
        \centering
	\includegraphics[width=0.6\linewidth]{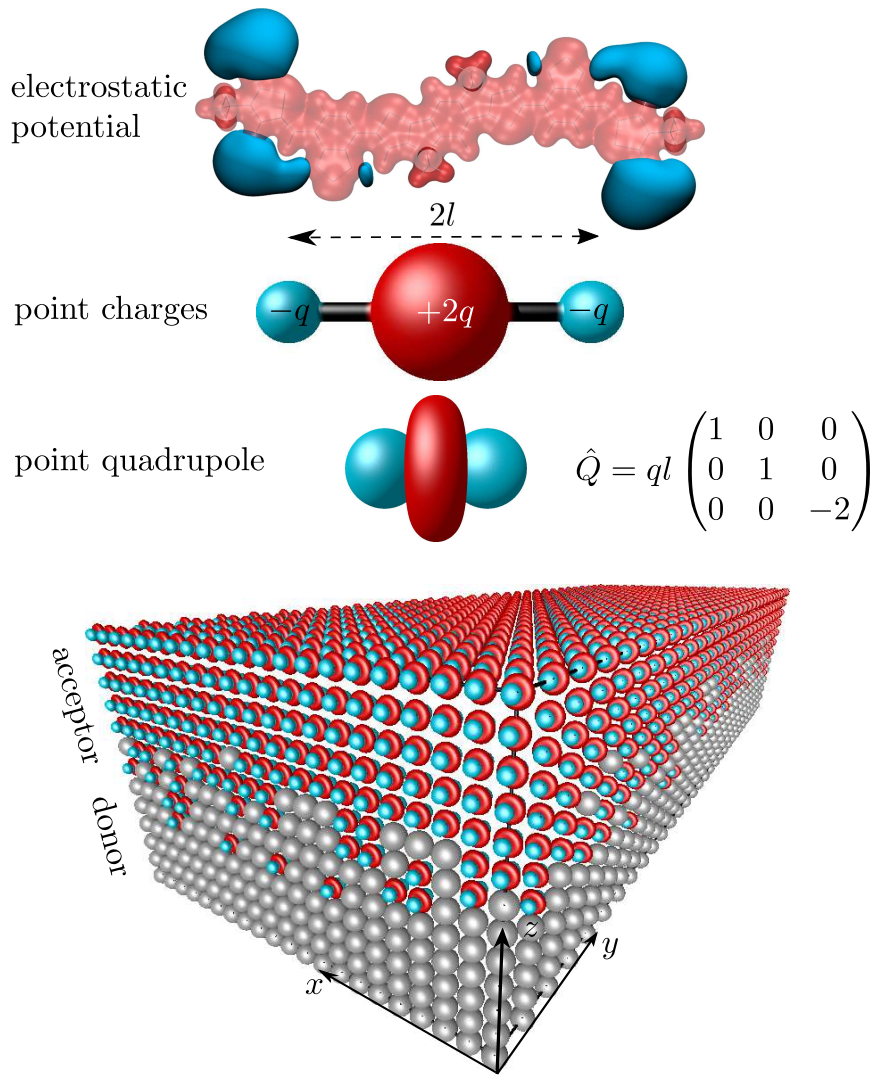}
	\caption{Isopotential surfaces at $-0.025$ V (blue) and red $+0.05$ V (red) of an represenative NFA molecule, calculated using DFT at B3LYP 6-311g(d,p) level of theory. Below, a set of point charges assigned to donor and acceptor units and an approximation of the molecular electrostatic potential with a potential of a linear quadrupole. Bottom:  Lattice model of the donor-acceptor interface. The donor phase (gray spheres) is electrostatically neutral, while the acceptor phase is approximated by a lattice of point quadrupoles. $\hat{Q}$ is a quadrupole tensor oriented along $y$-axis and $q$ is a point charge.}
    \label{fig:lattice}
\end{figure}

We begin by showing that the A-D-A molecular architecture of the acceptor is essential for the NFA design, because it lowers the barrier for the dissociation of the CT state into the CS state. To show that the A-D-A molecular architecture provides the required increase of the CT state energy, we adopt a simple model, leaving involved atomistic-level calculations to section~\ref{sec:photovoltaic}. Most NFAs have zero dipole moments, or dimerize in a unit cell such that the dipoles of neighbors compensate each other~\cite{perdigon-toro_barrierless_2020}. We therefore reduce the electrostatics representation of a NFA molecule to a non-polarizable linear quadrupole, as depicted in {\bf Figure~\ref{fig:lattice}}. Note that the A-D-A molecular architecture corresponds to {\em positive} values of $q$, while D-A-D would result in negative $q$. 

We cut the donor-acceptor interface out of a lattice of quadrupoles, as depicted in {\bf Figure~\ref{fig:lattice}}. Lattice parameters, chosen to match typical NFA crystals, and the geometry of the  interface are described in {\bf Supplementary Note~\ref{si:lattice}}. We then place an electron, modeled as a negative point charge, on every acceptor molecule and evaluate its electrostatic interaction energy, $\text{EA}^\text{A}_\text{elec}$ with the surrounding quadrupoles either explicitly, as described in {\bf Supplementary Note~\ref{si:lattice}}, or using the aperiodic Ewald summation technique~\cite{poelking_long-range_2016, davino_electrostatic_2016}. $\text{EA}^\text{A}_\text{elec}$ is, in fact, the solid-state contribution to the electron affinity, 
\begin{align}
\text{EA}^\text{A}(\bm r) = \text{EA}^\text{A}_\text{gas} + \text{EA}^\text{A}_\text{elec}(\bm r),
\end{align}
which explicitly depends on the spatial coordinate $\bm r$ because of the concentration gradient of the acceptor in the interfacial region. 

\begin{figure}[ht]
        \centering
	\includegraphics[width=0.5\linewidth]{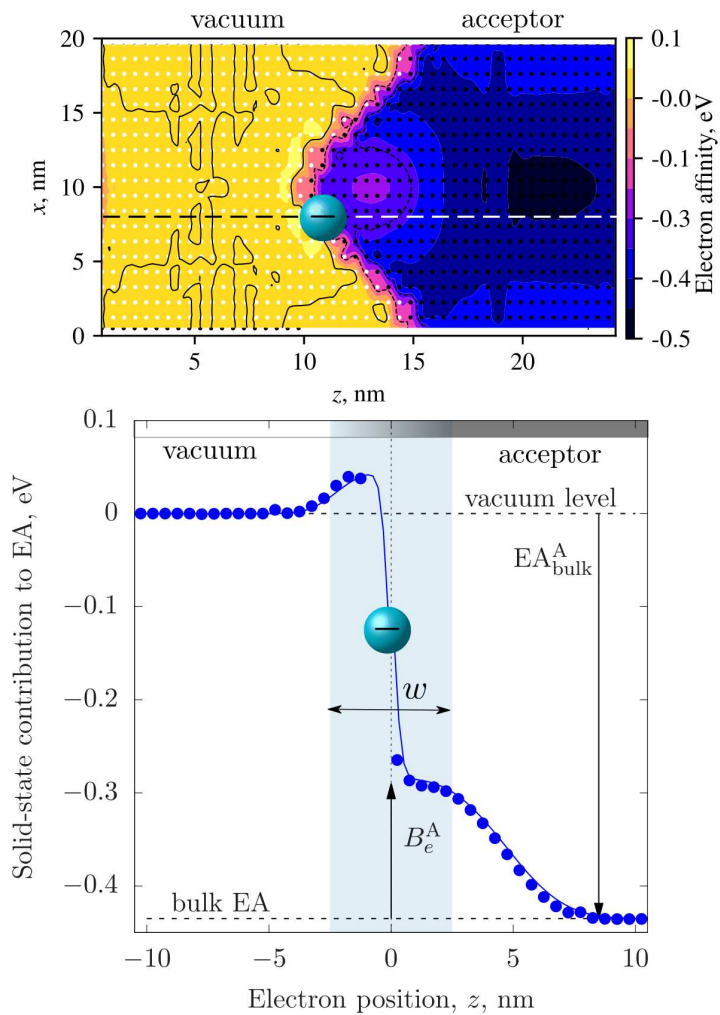}
	\caption{ Top: Solid-state contribution to the electrostatic potential, $xz$ cross-section. Bottom: $x$ profiles along the cross-section at $x=7.5 \text{nm}$. The dashed line shows the center of the intermixed interface and blue area shows the width of the intermixed region. The lattice parameters were estimated from IEICO crystal structure: $a=2$ nm, $b=1$ nm, $c=0.4$ nm, the maximum value of the quadrupole tensor is oriented along the $y$ axis ($Q_{20}= 5$ $ea_0^2$,  $Q_{22c}=-15/\sqrt{3}$ $ea_0^2$).} 
	\label{fig:dissociation}
\end{figure}

This dependence is shown in {\bf Figure~\ref{fig:dissociation}} for the acceptor-vacuum interface of width $w = 5\, \text{nm}$. The cross-section of the $xz$ map shows a gradual change of electron affinity from the bulk to the interfacial value, $\text{EA}^\text{A}_\text{int} = \text{EA}^\text{A}_\text{bulk} + B_e^\text{A}$, which can be approximated by an empirical function, $\text{EA}^\text{A}(z)  = \text{EA}^\text{A}_\text{bulk} + B_e^\text{A} \exp \left[ - (|z|/w)^3 \right]$, as shown in {\bf Figure~\ref{fig:dissociation}} by the solid line.

The introduced bias potential, $B_e^\text{A}$, is the manifestation of the charge-quadrupole interactions and the interfacial concentration gradient. As such, $B_e^\text{A}$, depends on the molecular packing, interface width $w$, and quadrupolar tensor of the acceptor. In {\bf Supplementary Note~\ref{si:unticell}} we show that the main contribution to the bias potential is due to the shortest intermolecular distance, that is the $\pi-\pi$ stacking distance of NFAs. In {\bf Supplementary note~\ref{si:width}} we conclude that we always need a certain degree of intermixing to observe energy level bending. In fact, the value of the bias potential increases gradually with the degree of intermixing, saturating for 5-8 nm thick interfaces. In {\bf Supplementary Note~\ref{si:orientation}} we also show that the bias potential is very sensitive to the molecular orientation at the interface, with the parallel and perpendicular orientations favoring CT state dissociation. 

It is impossible to evaluate the bias potential without explicit knowledge of molecular packing and intermixing at the interface. However, a qualitative comparison of bias potentials of different acceptors is still possible: in {\bf Supplementary Note~\ref{si:proportionality}} we show that the bias potential is (approximately) proportional to the solid-state contribution to EA, $B_e^\text{A}  \simeq - \xi_e  \text{EA}_\text{elec}^\text{A}$, where $\xi_e$ depends on the molecular packing at the interface and its shape. Therefore, one can compare the solid state contributions in thin NFA films instead of explicit calculations of interfacial electrostatic potentials. We will make use of this observation when discussing atomistic models of NFAs in section~\ref{sec:photovoltaic}.

\section{Energy level diagram}

\begin{figure}[ht]
        \centering
	\includegraphics[width=0.6\linewidth]{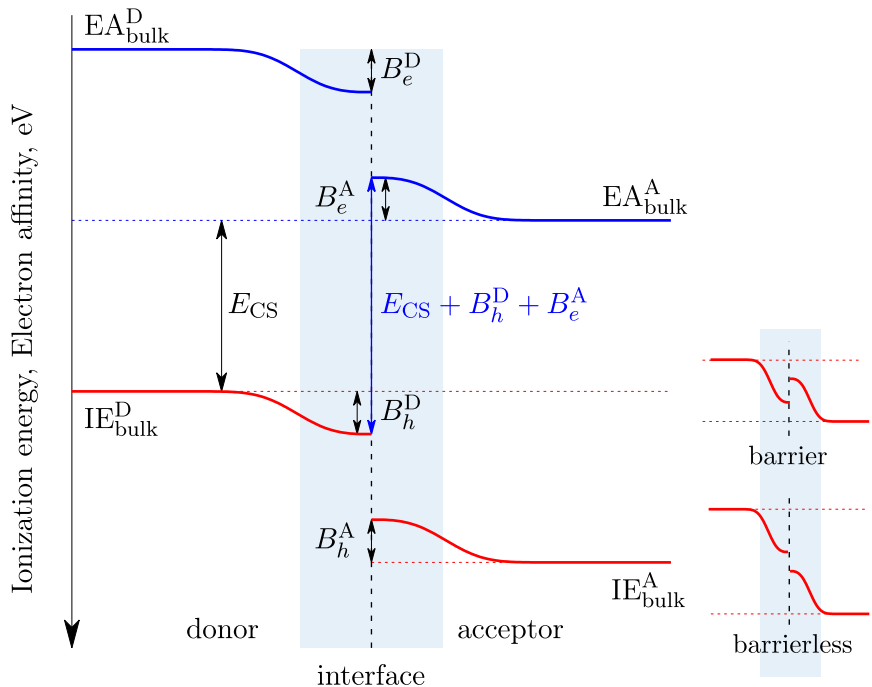}
	\caption{ Sketch of the energy level diagram at the donor-acceptor interface illustrating the concepts of the energy level bending and (positive) interfacial bias potential. The electron is more stable in the phase with lower electron affinity (larger negative energy values) and the hole is more stable in the material with the higher ionization energy. The direction of the energy level bending corresponds to the A-D-A molecular architecture and long molecular axes oriented parallel to the donor-acceptor interface. Positive interfacial bias destabilizes the charge transfer state, helping to dissociate it into the charge separated state. The inset shows energy diagrams for hole dissociation with and without energy barrier. Note that here we assumed that the donor phase also provides bend bending that destabilizes the CT state. 
	}
	\label{fig:oscdiagram}
\end{figure}

We now incorporate the energy level bending into the conventional energy level diagram of an organic solar cell. This diagram shows a relative alignment of ionization energies and electron affinities of the donor and acceptor and provides a convenient way of illustrating the  donor-acceptor character of the interface. {\bf Figure~\ref{fig:oscdiagram}} shows the energy level diagram with the level bending at the donor-acceptor interface and provides a useful insight into the barrier of dissociation of the excited and charge transfer states. According to this diagram, excitation energies of the donor and acceptor are
\begin{align}
E^{\text{A}^*} &= \text{EA}^\text{A}_\text{bulk} - \text{IE}^\text{A}_\text{bulk} - E^{\text{A}^*}_\text{binding} \;, \\
\nonumber
E^{\text{D}^*} &= \text{EA}^\text{D}_\text{bulk} - \text{IE}^\text{D}_\text{bulk} - E^{\text{D}^*}_\text{binding} \;,
\end{align}
where $E^{\text{A, D}^*}_\text{binding}$ are the binding interaction energies of the electron and the hole in the excited acceptor/donor, in the solid state. 

Due to the energy level bending, the energy of the charge transfer state is increased by a positive bias potential, $B$, 
\begin{align}
\nonumber
E^{\text{CT}} &= \text{EA}^\text{A}_\text{interface} - \text{IE}^\text{D}_\text{interface} - E^{\text{CT}}_\text{binding} = \\ 
 &= \text{EA}^\text{A}_\text{bulk} - \text{IE}^\text{D}_\text{bulk} - E^{\text{CT}}_\text{binding} + B \;, 
\end{align}
where $E^{\text{CT}}_\text{binding}$ is the binding energy of the electron and the hole in the CT state, $B = B_e^\text{A} + B_h^\text{D}$. The energy level bending leads to an electron destabilization in the acceptor, and hole destabilization in the donor, by amounts $B_e^\text{A}$ and $B_h^\text{D}$, respectively. Note that both the donor and the acceptor contribute to the level bending. For the donor, the simultaneous stabilization of an electron and a destabilization of a hole at the interface corresponds to what the D-A-D architecture of NFA would provide. 

Finally, the energy of the charge separated state,
\begin{align}
 E^\text{CS} = \text{EA}^\text{A}_\text{bulk} - \text{IE}^\text{D}_\text{bulk} \;,
\end{align}
depends only on the cation and anion energies in the bulk of the film.

With these expressions we are ready to analyze the driving forces of the relevant for OSC transitions, namely $\text{A}^*$ and $\text{D}^*$ to the CT state as well as the CT to CS state transition. 

\section{Formation of the charge transfer state}

In a bulk heterojunction solar cell, both donor and acceptor excitons can contribute to the pool of CT states. In solar cells with NFAs, the larger optical gap of the donor, as compared to the acceptor, leads to a fast resonant energy transfer from the donor to acceptor~\cite{zhong_sub-picosecond_2020, karuthedath_intrinsic_2021}. The donor-to-acceptor energy transfer is long-range and is faster than the electron transfer which relies on many charge transfer reactions. Hence, the solar sell efficiency is primarily dependent on the hole transfer from the excited acceptor to the CT state~\cite{moore_ultrafast_2020}.

The driving force for the hole transfer, or $\text{A}^* \rightarrow \text{CT}$ reaction, has three distinct contributions,
\begin{align}
 \Delta E_{ \text{A}^* \rightarrow \text{CT} } &= E^{\text{A}^*} - E^{\text{CT}} = \Delta \text{IE} - B - \Delta E_\text{binding} \, ,
\end{align}
where $\Delta E_\text{binding} = E^{\text{A}^*}_\text{binding} - E^{\text{CT}}_\text{binding}$ is the reduction in the binding energy when going from an excited acceptor to the charge transfer state. 

It might seem that the electron-hole binding is much stronger in the excited state than in the CT state because the electron and the hole are, on average, further apart in the CT state. This is true if only the classic Coulomb interaction between the electron and hole are considered. The repulsive exchange interaction decays much faster than the attractive Coulomb interaction as distance increases, which results in smaller binding energy of CT state than that of excited state. Furthermore, in a solid state, the induction stabilization of two practically independent charges forming the CT state is much larger than the stabilization of a dipole (or even a quadrupole) of an excited state. In other words, the electron-hole dissociation is impeded by the Coulomb energy but promoted by the gain in the induction energy. In fact, calculations based on polarizable force-fields (see {\bf Supplementary Note~\ref{si:stabilization}}) predict that in NFAs $\Delta E_\text{binding} \sim 0.3 - 0.5$~eV, depending on the compound.

\begin{figure}[ht]
        \centering
	\includegraphics[width=0.6\linewidth]{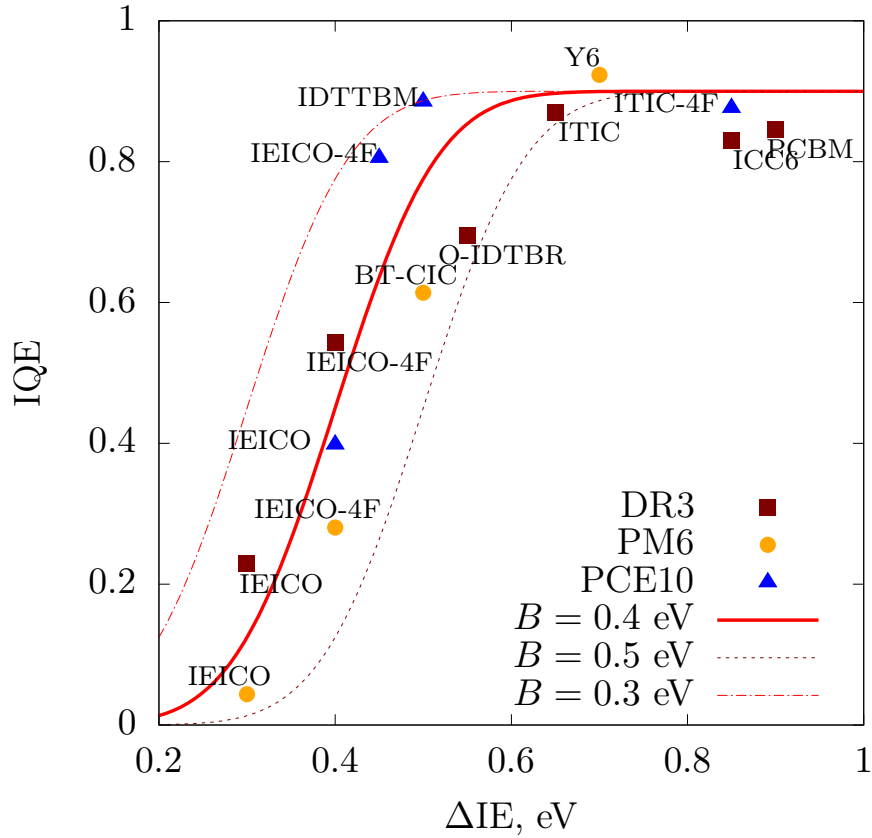}
	\caption{Internal quantum efficiency of optimized bulk heterojunction donor-NFA devices as a function of the IE offset between donor and acceptor for the small molecule donorfor DR3 (squares), PM6 (circles), and PCE10 (triangles). IE determined by ultraviolet photoelectron spectroscopy.}
	\label{fig:iqe}	
\end{figure}

The main driving force is therefore the offset of the ionization energies at the interface, or ionization eneries in the bulk reduced by the bias potential. This conclusion can be validated by examining the charge dissociation efficiency as a function of the offset of the donor-acceptor ionization energy. Since IQE is directly proportional to the fraction of excited to CT state transitions, and assuming that only the {\em barrierless} transitions, i.~e., those with positive $\Delta E_{ \text{A}^* \rightarrow \text{CT} }$ (see the inset of Figure~\ref{fig:oscdiagram}) contribute to the formation of CT states, we obtain 
\begin{align}
\text{IQE} = \text{IQE}^\text{max} \text{erfc} (\bar{B}-\Delta \text{IE} / \sigma) \, ,
\label{eq:IQEfit}
\end{align}
where we assumed that the rough donor-acceptor interface leads to a Gaussian-distributed bias potential of variance $\sigma^2$.

IQEs of more than 20 donor-acceptor combinations are shown in {\bf Figure~\ref{fig:iqe}}, together with the fit function, Eq.~\ref{eq:IQEfit}. Remarkably, for all donors we obtain $\bar{B} \sim 0.4$ eV, which indicates that this is the optimal value of the {\it mean} bias, since all solar cells are highly optimized. In the next section we discuss the origin of this universal behavior. As a consequence, an offset of at least $\Delta \text{IE} \ge \bar{B} + \sigma \approx 0.5$ eV is needed to reach maximum IQE, which is a clear design rule for choosing appropriate donor-acceptor combinations~\cite{karuthedath_intrinsic_2021}.

\section{Dissociation of the charge transfer state}
\label{sec:ctdissoc}
\begin{figure}[ht]
        \centering
	\includegraphics[width=0.6\linewidth]{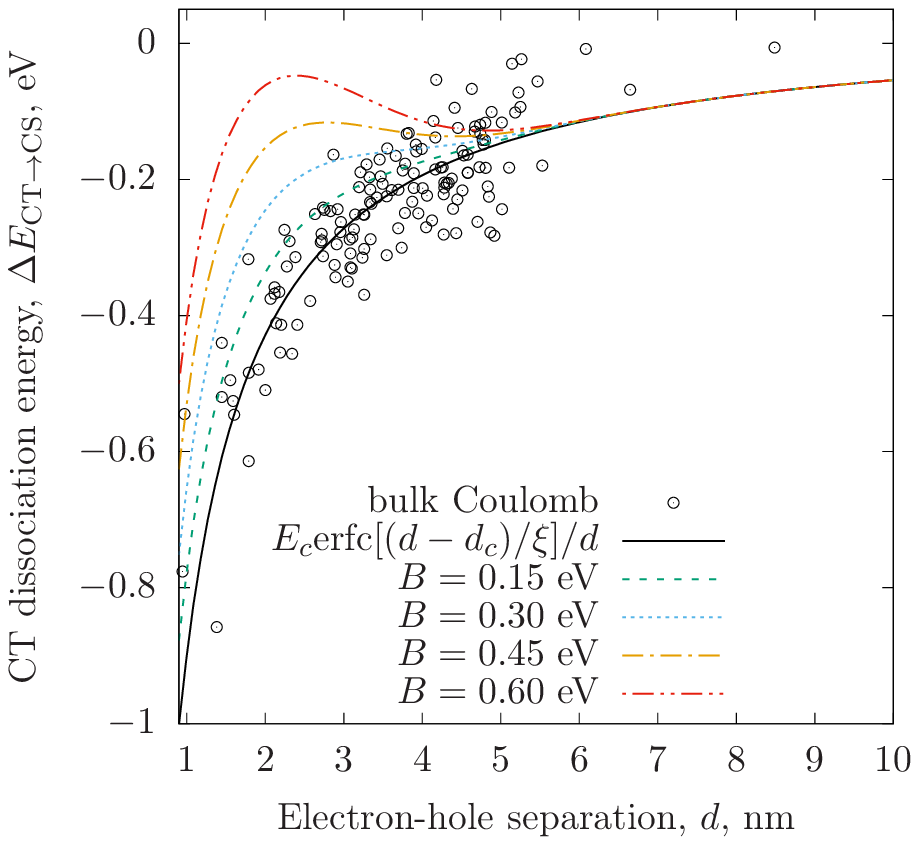}
	\caption{ Energy bending of CT state for different values of bias potential $B =0.3 - 0.5~\text{eV}$. A dashed line shows energy of a CT state without electrostatic effect associated with an interface.} 
	\label{fig:coulomb}	
\end{figure}

To dissociate a CT state, an electron (or a hole) needs to overcome the CT state dissociation energy, or the energy difference between the CT and CS states, which amounts to the Coulomb binding energy of the CT state in the solid state, reduced by the bias potential,
\begin{align}
\Delta E_{\text{CT} \rightarrow \text{CS} }=E^{\text{CS}}-E^{\text{CT}}=E^{\text{CT}}_{\text{binding}}-B \, .
\label{eq:ct2cs}
\end{align}

From this expression it is clear that the barrierless transition occurs when $B = E^{\text{CT}}_{\text{Coulomb}} \sim 0.5$~eV. In other words, the optimal value of the bias potential equals the Coulomb binding energy of an electron and hole at an ideally flat donor-acceptor interface. Notably, there is a remarkable agreement with the fit obtained during the analysis of the charge transfer state formation shown in Figure~\ref{fig:iqe}, indicating that for well-optimized NFA -based solar cells $B \sim 0.5$~eV.

One can further examine the CT state binding energy as a function of the electron-hole separation, $d$. {\bf Figure~\ref{fig:coulomb}} (symbols) shows interaction energy, calculated using polarizable force fields, as a function of electron-hole separation (see {\bf Supplementary Note~\ref{si:stabilization}} for computational details). In the absence of electrostatic bias, electron-hole interaction energy is set by the Coulomb attraction. At large separations it can be approximated by the screened Coulomb potential with a relative dielectric constant of 4. At short separations, simple dielectric screening is not applicable, and the binding energy can be as high as 0.4-0.5 eV~\cite{zhu_exciton_2018}. 
A phenomenological fit with $E^\text{CT}_\text{binding}(d) = E_c \text{erfc} \left( [{(d-d_c)/\xi}] \right)/d$ (solid line in {\bf Figure~\ref{fig:coulomb}}), combined with the electrostatic interfacial bias, $B(d) = B \exp \left( - [(d+d_0)/w]^3 \right) $, is shown for $w$ = 2 nm, $d_0 = 1$~nm and a range of bias potentials. One can see that for $B< 0.5$ eV dissociation energies are too high and the CT state dissociation cannot be  activated thermally~\cite{saladina_charge_2021}. We can therefore conclude that bias potentials on the order of 0.5 eV are required to dissociate CT state, in line with the results reported for one of the best performing NFA, Y6~\cite{perdigon-toro_barrierless_2020}.

\section{Photovoltaic characteristics}  
\label{sec:photovoltaic}

We will now correlate the bias potential to photovoltaic characteristics, namely the open circuit voltage, $V_\text{oc}$, and the short circuit current, $J_\text{sc}$. 

As we have seein, the lattice model of a rough interface provides a qualitative understanding of the energetics driving exciton dissociation into a pair of free charges. For compound pre-screening, it would be useful to be able to compare interfacial biases of different NFAs, without performing time-consuming simulations of the donor-acceptor interface. To do this, we can make use of the fact that the bias potential of a hole is proportional to the difference between the hole’s electrostatic energy in the donor and acceptor phases, see {\bf Supplementary Note~\ref{si:proportionality}}. Crystal fields of pristine layers of these compounds were evaluated by treating electrostatic and induction effects as the first- and second-order perturbations to the gas-phase energies~\cite{stone_distributed_2005,poelking_design_2015,poelking_impact_2015,poelking_long-range_2016}, as described in {\bf Supplementary Note~\ref{sec:ie_ea}}. 

\begin{figure}[ht]
        \centering
	\includegraphics[width=0.8\linewidth]{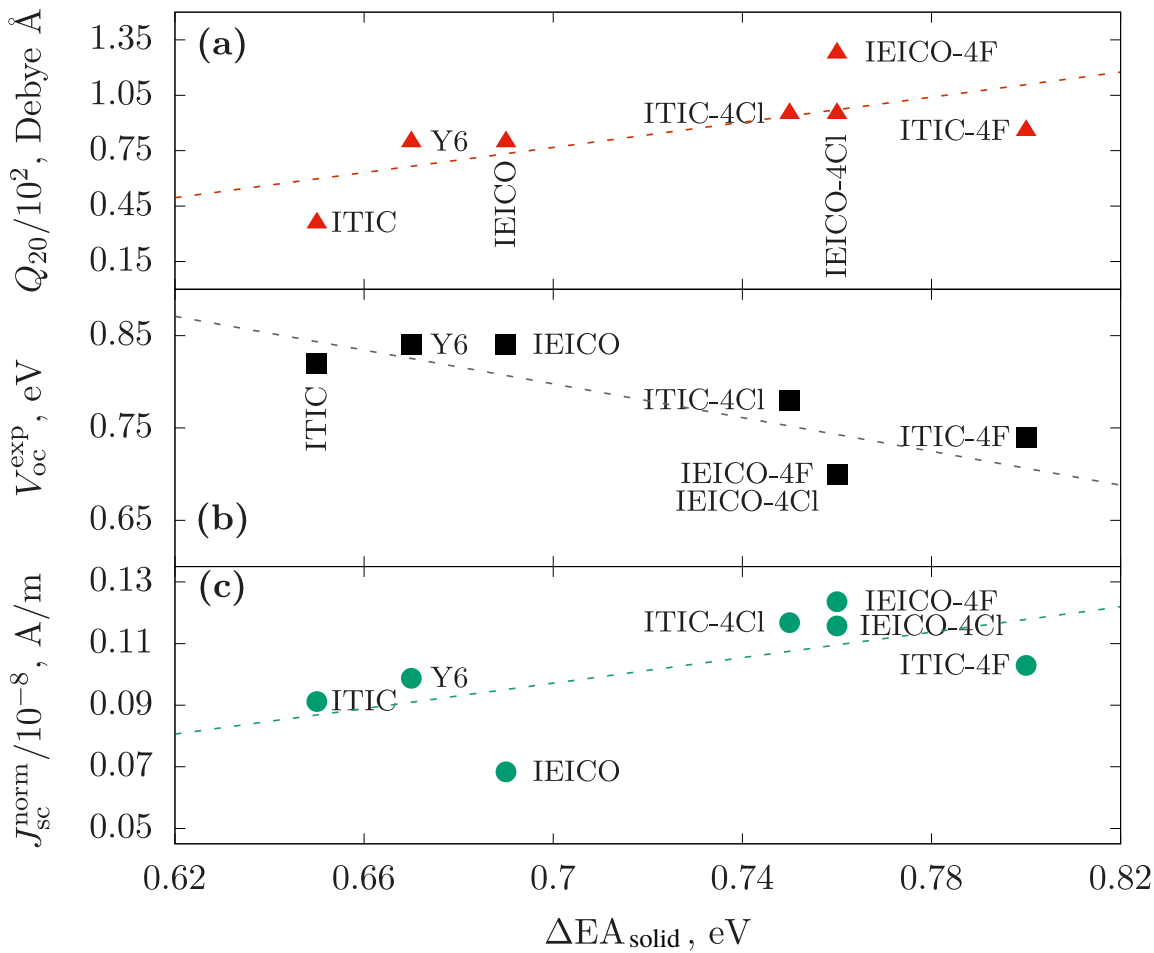} 
	\caption{Calculated quadrupole moment $Q_\pi$, given in atomic units. The actual values are summarized in {\bf Supplementary Note~\ref{sec:simulation}}. 	
	(a), experimental values of normalized by absorption current density $J^\text{norm}_\text{sc}$ (b), and open circuit voltage $V_\text{oc}^\text{exp}$ (c) versus calculated bias potential $B$. Solar cells were prepared and characterized as described in the {\bf Supplementary Note~\ref{si:experiment}}.}
	\label{fig:qvj}
\end{figure}

With the values of the bias potential at hand, we can now re-examine one of the key results of the lattice model, stating that the interfacial bias is proportional to the molecular quadrupole moment. Figure \ref{fig:qvj}(a) shows a correlation between the bias potential and the $Q_\pi$ quadrupole moment (perpendicular to the $\pi$-system) for all studied NFAs. In spite of differing crystal structures, molecular packings, and charge distributions, one can clearly see that these quantities correlate. This is an important conclusion for the overall chemical design, since the $Q_\pi$ component of the molecular quadrupole tensor can be easily evaluated and used to pre-screen NFAs. Using $Q_\pi$ as a descriptor will allow for fast screening of new compounds for organic solar cells, thereby avoiding complex parameters such as bias potential, $B$.

Since both the bias potential and the solid state contribution to the ionization potential are related to each other, we can anticipate a correlation between $V_\text{oc}$ and the bias potential. Indeed, as shown in Figure \ref{fig:qvj}(b), there is an inverse correlation between $B$ and $V_\text{oc}$. In other words, an increase in the bias potential leads to a decrease in voltage. This can be rationalized, in part, by examining the energy level diagram, as shown in {\bf Figure~\ref{fig:diagramlev}}: larger $\delta_e$ and $\delta_h$ lead to smaller $V_{oc}$, in the same way as larger crystal fields reduce the transport gap in the solid state. Apart from this obvious correlation, the molecular design itself favours the same behaviour: stronger acceptors in the ADA molecular architecture reduce the HOMO-LUMO gap of the molecule, but at the same time increase the charge flow within the molecule, leading to a larger molecular quadrupole moment (and, therefore, larger bias potential).

A different trend is observed for the short circuit current, $J_\text{sc}$. A correlation between $J_\text{sc}$ and the bias potential is shown in Figure \ref{fig:qvj}(c). To factor out the difference in currents due to 
the difference in absorption, we normalize $J_\text{sc}$ by the total absorbtion. Here, larger biases lead to stronger currents. Therefore, charge splitting is more efficient in systems with large biases, the conclusion we have already made in Section II by showing that the CT state dissociation barrier decreases with the increasing bias.
 
We can therefore conclude that the electrostatic crystal field leads to two competing effects. First, large $B$ favors efficient charge splitting but can also result in additional $V_\text{oc}$ losses. 
The trade-off can be estimated by analyzing quadrupolar moments of donor-acceptor combinations with high IQE, in particular the halogented ITIC series. The typical solid-state contribution ($x$ axis in Figure~\ref{fig:qvj}) of these NFAs varies in the 0.7-0.8 eV range, which, according to the Figure~\ref{fig:qvj}(a), corresponds to roughly $Q_{\pi}^\text{opt} \sim 100 ea_0^2 \approx 75 \, \text{Debye \AA}$. 

In principle, different donors could have a different contribution to the bias potential. This would manifest itself in different mean values of the bias potential $B$ in Figure 5. What we observe, however, is $\bar{B} \sim 0.4$ eV for all three -- chemically very different -- donors. This indicates that either all donors are already optimized in terms of their contribution to the intrefacial electrostatic field, or their role is secondary as compared to the acceptor molecules. 

\section{Computational screening}  
\label{sec:vscreen}

\begin{figure}[ht!]
         \centering
	\includegraphics[width=0.8\linewidth]{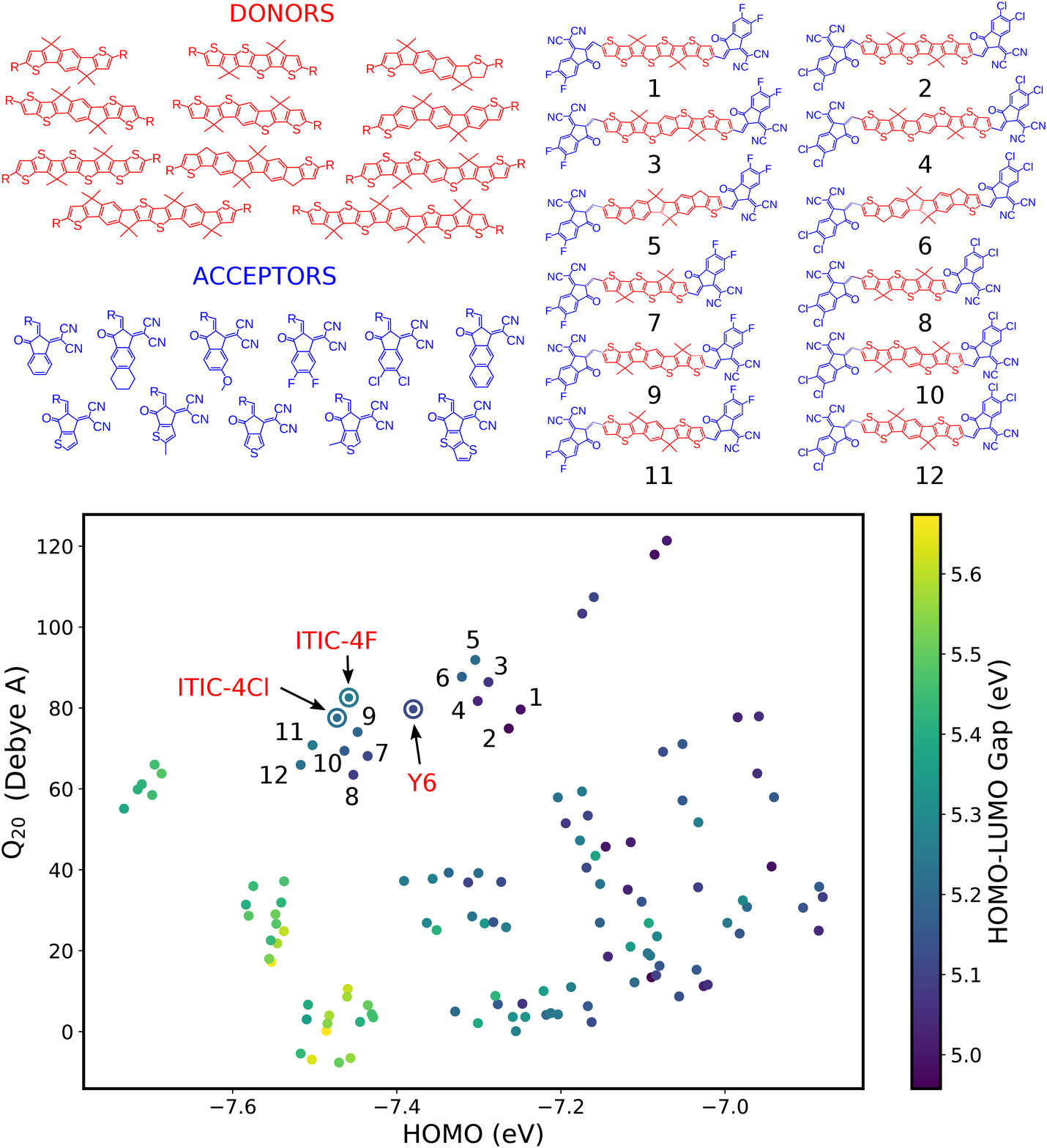}
 	\caption{
                Left: Chemical structures of acceptors and donors with $R$ depicting the donor-acceptor bonds. Right: Molecular structures of 12 selected compounds. Bottom: $Q_{20}$-HOMO plot for 121 A-D-A compounds as well as high-performance NFAs ITIC-4F, ITIC-4Cl, and Y6. Each point is colored according to its corresponding HOMO-LUMO gap value.
                }
 	\label{fig:screening}
 \end{figure}

To illustrate the practicality of the proposed design criteria, we performed a computational screening on a 121 molecules dataset constructed following the A-D-A design principle with the 11 donors and 11 acceptors shown in Figure~\ref{fig:screening} 1. All donors have inversion symmetry to ensure a small molecular dipole moment, which should facilitate the hole transport of the A-D-A compound~\cite{mondal_molecular_2021}. The acceptors are 2-methylene-(3-(1,1-dicyanomethylene)indanone) derivatives that have been used in previous studies~\cite{shen_nonfullerene_2018}.

As a pre-screening step, we computed $Q_{20}$, energies of the highest and lowest occupied molecular orbitals (HOMO and LUMO) for each compound. Computational details are provided in {\bf Supplementary Note~\ref{sec:simulation}}. The $Q_{20}$ versus HOMO plot is shown in Figure~\ref{fig:screening} 1, together with the values for high-performance NFAs such as ITIC-4F, ITIC-4Cl and Y6. It is clear form Figure 1 that 12 out of the 121 compounds cluster with efficient NFAs such as ITIC-4F, ITIC-4Cl, and Y6. These 12 compounds are also shown in Figure~\ref{fig:screening} 1. Not surprisingly, the compounds within this cluster share the same set of acceptor building blocks as ITIC-4F and ITIC-4Cl. In addition, compounds 11 and 12 are identical to ITIC-4F and ITIC-4Cl, except for the  side chains on the donor. This implies that further fine-tuning can be achieved by side-chain engineering.

We further compute the gas-phase IE, EA and energy of the S$_1$ state for the 12 selected compounds. Judging from these properties, which are summarized in {\bf Supplementary Note~\ref{sec:simulation}}, all 12 compounds are potential NFA candidates for efficient solar cells with PCE10 as a donor. A more rigorous tests could be done by performing solid-state corrections to the gas-phase properties, but this is computationally demanding as it requires predictions of the model morphologies. 

Overall, the proposed chemical design rules help to reduce the database of 121 compounds, constructed using 11 donors and 11 acceptors, to 12 potential candidates, which is an order of magnitude reduction for the compound pre-screening workflow. In fact, ten out these twelve compounds have been reported in the literature, and eight out of the ten resulted in solar cells of 10 to 15 \% efficiency: (1) F-IXIC with benzene side chains, $\sim 10$ \% ~\cite{shi_two_2020} ; (2) IXIC-4Cl with benzene side chains, 15 \% ~\cite{liu_nonfullerene_2019}; (3,4) patented~\cite{mitchell_organic_2019}; (5) ZITI-C with 2-butyloctyl side chains, $\sim 13$ \%~\cite{zhang_revealing_2019}; (7) 4TIC-C8-2F with benzene side chains, $\sim 11$ \% ~\cite{chen_improving_2021}; (9) BDCPDT-FIC with benzene side chains, $\sim 8$ \%~\cite{chang_new_2018} or FBDIC with benzene side chains, $\sim 12$ \%~\cite{cai_comparison_2019}; (10) BT-CIC with benzene side chains, $\sim 11$ \%~\cite{li_high_2017}; (11) ITIC-4F with benzene side chains, $\sim 13$ \%~\cite{jiang_employing_2019}; (12) ITIC-4Cl with benzene side chains, $\sim 14$ \%~\cite{zheng_over_2021}. 
This is a clear illustration of robustness and usefulness of the proposed screening.

\section{Conclusions}

We can now summarize the known and new chemical design rules that can help to narrow the search of efficient donor -- acceptor combinations for organic solar cells. 

Several observations can readily be made even without referring to the variation of the electrostatic potential at the donor-acceptor interface. For example, rigid elongated planar cores of NFAs favor the formation of spatially extended domains, about 10-30 nm in size~\cite{baran_reducing_2017}. Acceptor molecules are well-aligned within these domains, which leads to a narrow distribution of electron affinities, with half-widths on the order of  0.1 eV, facilitating good electron mobilities. Furthemore, electron affinities lower than -3 eV ensure trap-free electron transport~\cite{kotadiya_window_2019}. In fact, NFAs often exhibit ambipolar transport~\cite{kotadiya_window_2019, paterson_n-doping_2021} as their ionization energies are normally above 6 eV. Rigid planar cores and large electronic couplings result in superior exciton diffusion lengths, up to 50 nm~\cite{firdaus_long-range_2020}. Due to this, the bulk heterojunction becomes more robust with respect to the domain size variation.

In addition to these design rules, the interfacial energetics imposes constraints onto the molecular architecture of the acceptor. Donor-acceptor intermixing at the donor-acceptor interface leads to the electrostatic potential bending at the interface. Acceptor-donor-acceptor molecular architecture ensures a negative component of the quadrupole moment tensor along the molecular axis~\cite{poelking_impact_2015, schwarze_impact_2019}, which is aligned with the donor-acceptor interface. The resulting electrostatic potential destabilizes the charge transfer state and drives its dissociation into free charges. Potential bending of around 0.5 eV compensates the electron-hole Coulomb binding energy, leading to barrier-less dissociation of the CT state in free charges~\cite{perdigon-toro_barrierless_2020}. The energy level bending reduces the driving force required for hole transfer into the acceptor to the donor, leading to the formation of charge transfer states. As a result, 0.5 eV offset between ionization energies of the donor and acceptor is required for efficient hole transfer reactions~\cite{karuthedath_intrinsic_2021}. 

The key result of this work is that the intricate energetics of the donor-acceptor interface can be traced backed to the molecular crystal field. Since the latter is related to the molecular quadrupole, the magnitude of energy level bending at the interface correlates with the molecular quadrupole moment. Therefore, the key difference between A-D-A and D-A-D molecular architectures is the sign of the quadrupolar moment. The same molecular alignment but with D-A-D architecture would lead to negative bias potential, stabilizing the charge transfer state, and creating inefficient solar cells. Moreover, as a rule of thumb, $Q_{\pi} \approx 100\, e a_0^2 $ (75 Debye {\AA}) provides a balance between efficient exciton dissociation and open circuit voltage losses.   

Using the proposed chemical design rules we show that a database of 121 compounds constructed using 11 donors and 11 acceptors can be reduced to 10 potential candidates, providing over an order of magnitude reduction for the compound pre-screening.

\medskip
\textbf{Acknowledgements} \par 
A.M. has received funding from the European Union’s Horizon 2020 research and innovation programme under the Marie Sklodowska-Curie grant agreement No 844655 (SMOLAC). This publication is based on work supported by the KAUST Office of Sponsored Research (OSR) under award nos. OSR-2018-CARF/CCF-3079 and OSR-CRG2018-3746. D.A. also acknowledges the KAUST PSE Division for hosting his sabbatical in the framework of the Division’s Visiting Faculty program. D.A. acknowledges funding by the Deutsche Forschungsgemeinschaft (DFG, German
Research Foundation) for financial support through the collaborative research centers TRR 146, SPP 2196, and grant number 460766640. We thank Kostas Daoulas, Leanne Paterson, and Naoimi Kinaret for fruitful discussions and proof-reading of the manuscript. 

\medskip

%
\bibliographystyle{MSP}


\bibliography{literature_short}



\end{document}